\documentclass{article}

\usepackage[a4paper,
            bindingoffset=0.2in,
            left=1in,
            right=1in,
            top=1in,
            bottom=1in,
            footskip=.25in]{geometry}
\usepackage[utf8]{inputenc}
\usepackage{xcolor}

\definecolor{seablue}{HTML}{4834d4}
\definecolor{red}{HTML}{eb4d4b}
\definecolor{lightred}{HTML}{ff7979}

\usepackage{amsmath}
\usepackage{algpseudocode}
\mathchardef\mhyphen="2D

\newcommand{\union}{\cup}
\newcommand{\afunc}[1]{\texttt{\color{seablue}{#1}}}
\newcommand{\naturalnumber}[0]{z}
\newcommand{\naturalnumberspace}[0]{\mathcal{Z}}

\newcommand{\realnumber}[0]{x}
\newcommand{\realnumberspace}[0]{\mathcal{R}}
\newcommand{\powerset}[1]{\mathcal{P}(#1)}
\newcommand{\setwhere}{~|~}
\newcommand{\subfn}{\afunc{sub}}
\newcommand{\set}[1]{\{#1\}}
\newcommand{\lit}[1]{{\color{red}{\texttt{#1}}}}
\newcommand{\tuple}[1]{\langle #1 \rangle}

\newcommand{\hwmodel}[0]{\afunc{H}}
\newcommand{\hwmodelspace}[0]{HwModels}

\newcommand{\conchwmodel}[0]{\afunc{CH}}
\newcommand{\conchwmodelspace}[0]{ConcHwModels}

\newcommand{\hwmetric}[0]{m}
\newcommand{\hwmetrics}[0]{M}

\newcommand{\hwspecialize}{\afunc{specialize}}

\newcommand{\techparam}[0]{tp}
\newcommand{\techparamset}[0]{TP}
\newcommand{\techparamspace}[0]{TechPars}

\newcommand{\archparam}[0]{ap}
\newcommand{\archparamset}[0]{AP}
\newcommand{\archparamspace}[0]{ArchPars}

\newcommand{\techassignset}[0]{TA}

\newcommand{\archassignset}[0]{AA}

\newcommand{\adexpr}[0]{e}
\newcommand{\adexprs}[0]{E}
\newcommand{\adexprspace}[0]{Exprs}

\newcommand{\computeclass}[0]{cc}
\newcommand{\computeclassset}[0]{CC}
\newcommand{\computeclassspace}[0]{CompCls}

\newcommand{\hwclass}[0]{c}
\newcommand{\hwclassspace}[0]{HwCls}

\newcommand{\memclass}[0]{mc}
\newcommand{\memclassset}[0]{MC}
\newcommand{\memclassspace}[0]{MemCls}

\newcommand{\hwmemmetric}[0]{mm}
\newcommand{\hwmemmetricspace}[0]{MemMetrics}

\newcommand{\hwcompmetric}[0]{cm}
\newcommand{\hwcompmetricspace}[0]{CompMetrics}

\newcommand{\computetechparamspace}{CompTechPars}
\newcommand{\memtechparamspace}{MemTechPars}
\newcommand{\computearchparamspace}{CompArchPars}
\newcommand{\memarchparamspace}{MemArchPars}

\newcommand{\workload}{w}
\newcommand{\workloadset}{W}
\newcommand{\workloadspace}{Workloads}
\newcommand{\vertex}{v}
\newcommand{\vertices}{V}
\newcommand{\vertexspace}{Verts}
\newcommand{\edge}{e}
\newcommand{\edges}{E}
\newcommand{\edgespace}{Edges}

\newcommand{\measurementspace}{Measurements}
\newcommand{\measurement}{q}
\newcommand{\perfestimate}{\afunc{P}}
\newcommand{\perfestimatespace}{PerfEstimate}

\newcommand{\dragon}{DRAGON}
\newcommand{\dopt}{DOpt}
\newcommand{\dsim}{DSim}
\newcommand{\dgen}{DGen}
\newcommand{\darchopt}{Dopt2}

\newcommand{\primtypespace}[0]{PrimitiveType}
\newcommand{\dmemlib}[0]{\afunc{memLib}}
\newcommand{\dmemlibspace}[0]{DevMemLib}
\newcommand{\dprimlib}[0]{\afunc{primLib}}
\newcommand{\dprimlibspace}[0]{DevPrimLib}

\newcommand{\acctempllib}[0]{\afunc{accTempls}}

\newcommand{\memassignfn}[0]{\afunc{memType}}
\newcommand{\memassignfnspace}[0]{memAssignFuncs}
\newcommand{\archspec}[0]{a}
\newcommand{\archspecspace}[0]{A}

\newcommand{\agexpr}[0]{xe}
\newcommand{\agexprs}[0]{XE}
\newcommand{\agexprspace}[0]{XExprs}

\algnewcommand\algorithmicmatch{\textbf{match}}
\algnewcommand\algorithmiccase{\textbf{case}}
\algnewcommand\algorithmicassert{\texttt{assert}}
\algdef{SE}[MATCH]{Match}{EndMatch}[1]{\algorithmicmatch\ #1\ \algorithmicdo}{\algorithmicend\ \algorithmicmatch}%
\algdef{SE}[CASE]{Case}{EndCase}[1]{\algorithmiccase\ #1}{\algorithmicend\ \algorithmiccase}%
\algtext*{EndMatch}%
\algtext*{EndCase}%

\newcommand{\memstate}{ms}
\newcommand{\computestate}{cs}
\newcommand{\vertexstate}{vs}
\newcommand{\vertexstatelist}{VS}

\newcommand{\addtfunc}[1]{\texttt{\color{red}{#1}}}

\newcommand{\softprefetch}{\addtfunc{prefetch}}
\newcommand{\softgetstats}{\afunc{getStats}}
\newcommand{\softsplitvertex}{\afunc{splitVertex}}
\newcommand{\softvertexstate}{\afunc{getVertexState}}
\newcommand{\softmapcompute}{\addtfunc{mapToCompute}}
\newcommand{\softmemalloc}{\addtfunc{memAlloc}}
\newcommand{\softmapmemacc}{\addtfunc{mapMemAcc}}
\newcommand{\softhasspace}{\afunc{hasSpace}}

\newcommand{\algmapvertex}{\textsc{mapVertex}}
\newcommand{\algprefetchvertex}{\textsc{prefetchVertex}}
\newcommand{\algmapworkload}{\textsc{mapWorkload}}

\newcommand{\memtype}[0]{mt}
\newcommand{\memtypespace}[0]{MemTypes}

\usepackage{algorithm}
\usepackage{mathtools}
\usepackage{amsmath}
\usepackage{amssymb}
\usepackage{pifont}
\newcommand{\cmark}{\ding{51}}%
\newcommand{\xmark}{\ding{55}}%
\usepackage{array}
\newcolumntype{P}[1]{>{\centering\arraybackslash}p{#1}}
\usepackage{color}
\usepackage{amsmath}
\usepackage{algorithm}

\title{\textbf{DRAGON (Differentiable Graph Execution)} : A suite of Hardware Simulation and Optimization tools for Modern Workloads}

\author{Khushal Sethi \thanks {khushal@stanford.edu}\\ Department of Electrical Engineering, Stanford University}

\begin{document}

\maketitle

\begin{abstract}

We introduce DRAGON, a fast and explainable hardware simulation and optimization toolchain that enables hardware architects to simulate hardware designs, and to optimize hardware designs to efficiently execute
workloads. 

The DRAGON toolchain provides the following tools: Hardware Model Generator (DGen), Hardware Simulator (DSim) and Hardware Optimizer (DOpt).

DSim provides the simulation of running algorithms (represented as data-flow graphs) on hardware described. DGen describes the hardware in detail, with user input architectures/technology (represented in a custom description language). A novel methodology of gradient descent from the simulation allows us optimize the hardware model (giving the directions for improvements in technology parameters and design parameters), provided by Dopt.

DRAGON framework (DSim) is much faster than previously avaible works for simulation, which is possible through performance-first code writing practices, mathematical formulas for common computing operations to avoid cycle-accurate simulation steps, efficient algorithms for mapping, and data-structure representations for hardware state.

DRAGON framework (Dopt) generates performance optimized architectures for both AI and Non-AI Workloads, and provides technology targets for improving these hardware designs to 100x-1000x better computing systems.

\end{abstract}










\section{Introduction}

21st-century computing systems are dramatically different from the 20th-century ones. Unlike traditional processors that dominated 20th-century computing, domain-specific accelerators are rising in the 21st century. The diversity of applications, algorithms and accelerator hardware architectures is changing very rapidly. For example, more than 200 hardware accelerators for AI inference and training have been published over the past 3-4 years. Beyond AI, hardware accelerators for data analytics, graph processing, genomics and security are also growing.

In addition, the computing demands of many of these applications are increasing at unprecedented rates. For example, the computing demands of AI workloads are doubling every 3-4 months. The latest GPT-3 by OpenAI requires 355 GPU-years to train. 

Major opportunities created by these 21st-century computing trends also bring their unique challenges:
Traditional processor simulators are no longer sufficient to create 21st century hardware architectures. Creating a different simulator for each application-accelerator configuration is infeasible. 

Supporting a new algorithm and architecture using current simulators requires significant development time (many person-months). By the time simulators are created, the accelerators they target often become obsolete. To keep up with the fast pace, new simulation frameworks are required. 

The performance (throughput, energy) demands of emerging applications cannot be met by architectural improvements or (incremental) technology improvements alone. While accelerators bring benefits over processors, further improvements (e.g., 100X Energy-Delay-Product (EDP) benefits) require significant technology improvements in conjunction with architecture improvements. This creates a critical need to translate application needs into technology targets. 

We propose a new, elegant, fast and explainable framework (Fig. 1) for end-to-end joint co-exploration and co-optimization of technologies, architectures and algorithms. Our framework overcomes the above challenges. 

The features of our framework are: 

\begin{itemize}

    \item Fast Performance Estimation (both energy and throughput) for a wide range of AI (inference and training) and non-AI workloads on many accelerator architectures.
    
    \item Technology Target derivation from workload performance objective (in terms of execution time, energy consumption etc.). It is a first of its kind feature which runs in seconds vs. traditional trial and error approaches that can be highly inaccurate (i.e., can miss important design points during technology space exploration) and slow (i.e., weeks or longer per exploration). 
    
    \item Accelerator architecture design optimization for those technology targets (e.g., in terms of systolic array structure, processing element dimensions, global buffer organization, connectivity to on-chip and off-chip memory for AI accelerators and the entire compute-memory subsystem for non-AI accelerators) that achieve desired system-level performance (e.g, Energy Efficiency, Execution Time).  Our framework provides accelerator design generation for a wide range of workloads, with designs supporting multiple data-flow graphs, all within a time frame of a few seconds. 
\end{itemize}

Our results are made possible by our unique approach which 
utilizes the (1) differentiable component (memory, compute, interconnects, connectivity) models from the technology parameter space created in the background, (2)
combines the created component models with workload data flow/control-data flow graphs through appropriate scheduling/mapping steps for fast performance estimation, and
(3) uses special (and provably correct) techniques to derive gradients of application performance metrics with respect to technology parameters.

The paper is organized as follows : Section 2 provides our methodology which creates data-flow graph and control-data graph representation for supporting multiple types of AI workloads, and Non-AI workloads respectively. User can specify different accelerator designs in a custom Architecture-description language (ADL) for the support of multiple data-flow AI/ control data-flow Non-AI architectures. The architecture description taken as input is used to create a hardware representation and a mapping process of the data-flow/control-data flow graph on the accelerator is used to generate the performance outputs (latency, energy consumption, resource utilization).

Section 3 provides the methodology that describes a Technology Description Language, and takes device-level technology parameters as input. Section 4 describes custom functions created to utilize device-level technology parameters and design parameters from ADL to create performance models (latency/energy etc.) of hardware components. Section 5 describes utilizing hardware component performance models to generate execution statistics for the entire application.


\section{Background}

A number of performance estimation simulators exist in computer architecture \cite{bashir2019power, sethi2021efficient, ji2020reconfigurable,ji2020compacc, sethi2020nv, sethi2020design, sethi2018low, sethi2019optimized}, which given an application can produce the performance estimates of its execution latency and energy consumption.

AI-specific simulators such as Scale-Sim \cite{samajdar2018scale}, Timeloop \cite{parashar2019timeloop}, NN-Dataflow \cite{yang2020interstellar}, Zig-Zag \cite{DBLP:journals/corr/abs-2007-11360}, NAAS \cite{lin2021naas}, have also been created which target a subset of CNN-specific workloads, but don't support modern AI workloads such as Transformers, Graph Neural Networks, Deep Recommendation Models, Generative Models etc. 

Running large AI workloads in these simulators is still a slow process, because the development of these simulators are not designed with runtime as a first priority. 

Further, these simulators cater to one or the other subset of architectures and don't support the wide-range of architectures we support.

Table \ref{tab:comparison} compares the capabilities of our framework with other open source available tools.  

\begin{table*}[!htbp]
\footnotesize
  \centering
\begin{tabular}{|P{1.3cm}|P{1.3cm}|P{1cm}|P{1cm}|P{1.4cm}|P{1.4cm}|P{1.3cm}|P{1.2cm}|P{1.2cm}|P{0.9cm}|}
\hline \textbf{Simulators} & \textbf{Perf. Est.} & \textbf{Mapping} & \textbf{CNNs} & \textbf{DLRMs/
Transformers/
GNNs} & \textbf{Non-AI Workloads} & \textbf{Algorithm Search} & \textbf{Derive Accel. Design} & \textbf{Derive Techn. Targets}. & \textbf{Run Time
(Avg.)} \\
\hline  \textbf{Scale-Sim (ARM) \cite{samajdar2018scale}} & \textbf{Latency}  &
\textbf{Single} & \cmark &
\xmark & \xmark & \xmark & \xmark & \xmark &
$\sim 10^3$ \\
\hline \textbf{Timeloop (Nvidia)}
& \textbf{Latency, Energy} &
\textbf{Single} & \cmark &
\xmark &
\xmark & \xmark & \xmark & \xmark &
$\sim 10^2$ \\
\hline \textbf{NN-Dataflow \cite{yang2020interstellar}} & \textbf{Latency} &
\textbf{Multiple (for maximal reuse)} &
\cmark &
\xmark &
\xmark &
\xmark &
\xmark & 
\xmark &
$\sim 5*10^2$ \\
\hline \textbf{Zig-Zag \cite{DBLP:journals/corr/abs-2007-11360}} & \textbf{Latency, Energy} & \textbf{Multiple} & \cmark & \xmark & \xmark &  \xmark & \textbf{Yes,
(Memory Design using Sweep)} & \xmark &
$\sim 10^4$ \\
\hline \textbf{NAAS\cite{lin2021naas}} & \textbf{Latency, Energy} & \textbf{Multiple} & \cmark & \xmark & \xmark &  \cmark &  \cmark & \xmark &
--\\
\hline \textbf{Ours} & \textbf{Latency, Energy} & \textbf{Multiple 
(w.r.t objective)} & \cmark & \cmark & \cmark & \cmark & \textbf{Yes, Single-Pass} & \textbf{Yes,
Single-Pass} & \textbf{1} \\
\hline 
\end{tabular}%
  \label{tab:comparison}%
  \caption{Comparison of our Work with Previously Published Results}
\end{table*}

\begin{figure}
    \centering
    \includegraphics[scale=0.3]{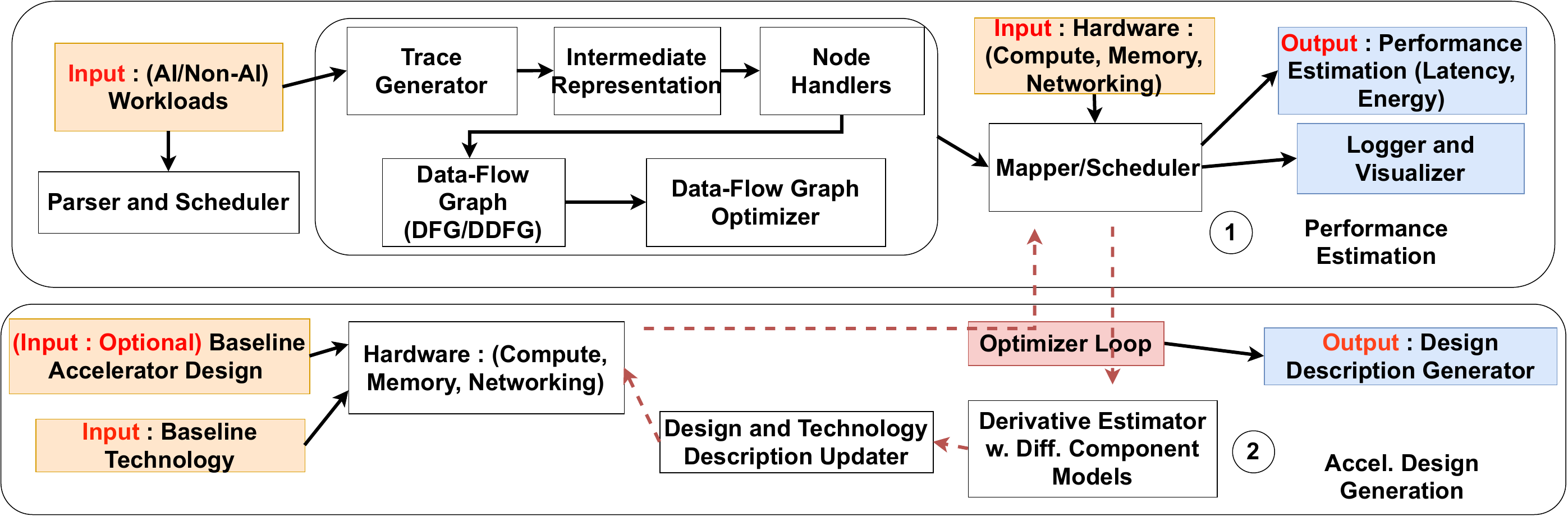}
    \caption{Overview of the Framework}
    \label{fig:my_label}
\end{figure}

\begin{figure}
    \centering
    \includegraphics[scale=0.40]{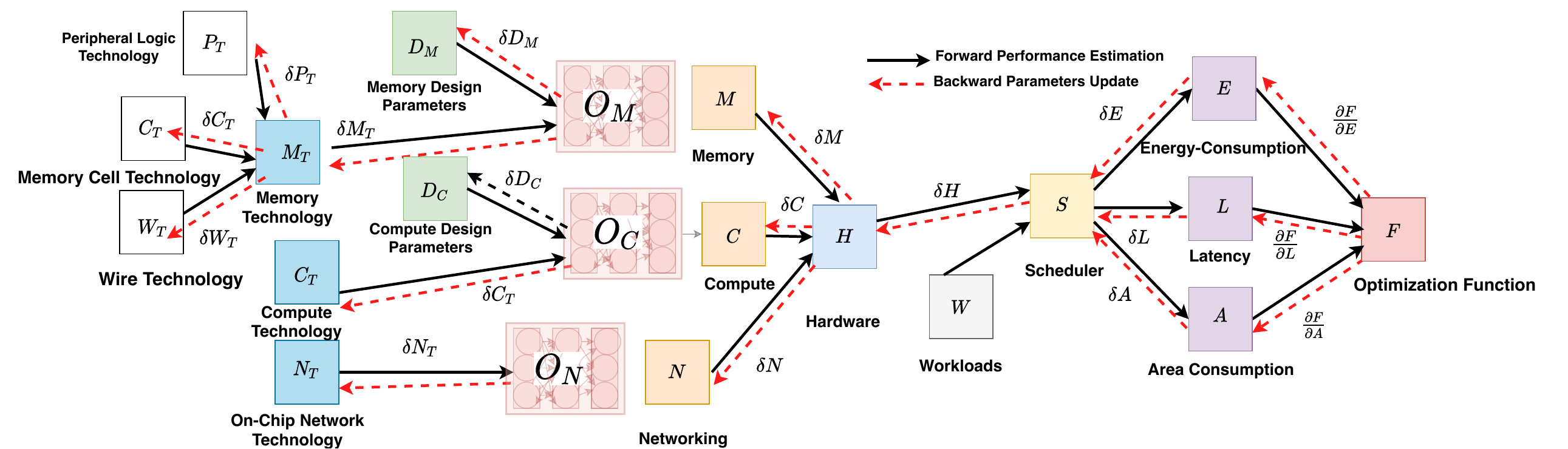}
    \caption{Forward and Backward Phase Execution}
    \label{fig:my_label}
\end{figure}
\section{The Hardware Model}

\begin{table*}[t]
\footnotesize
\begin{tabular}{cccc}
    \textbf{Parameter Type} & \textbf{Value} & \textbf{Classes} & \textbf{Parameter Names} \\
    \hline
     \textbf{\computetechparamspace} &  $\realnumberspace$ & $\computeclass \in \computeclassspace$ & $\lit{wireCap},\lit{wireResisist}$  \\
      &  $\naturalnumberspace$ && $\lit{node}$  \\

     \textbf{\memtechparamspace}& $\realnumberspace$& $\memclass \in \memclassspace$ & $\lit{wireCap},\lit{wireResist}$\\
      & & & $\lit{cellReadLatency}, \lit{cellAccessDevice}$\\
          & & & $\lit{cellReadPower}, \lit{cellLeakagePower}$, \lit{cellArea}\\
      & $\naturalnumberspace$ & &  $\lit{peripheralLogicNode}$\\
      \hline
      \textbf{\computearchparamspace} & $\naturalnumberspace$ & $\lit{systolicArray}$ & \lit{sysArrX},\lit{sysArrY}, \lit{sysArrN}\\\
       & & $\lit{vector}$ & $\lit{vectDataWidth}, \lit{vectN}$\\
       & & $\lit{macTree}$ & $\lit{mTreeX}, \lit{mTreeY}, \lit{mTreeTileX}, \lit{mTreeTileY}$\\
       & & $\lit{fpu}$ & $\lit{fpuN}$\\
       & & $\lit{SoC}$ & $\lit{frequency}$\\

      \textbf{\memarchparamspace} &  $\naturalnumberspace$ & $\memclass \in \memclassspace$ & $\lit{capacity},\lit{bankSize},\lit{nReadPorts}$\\
      \hline
          \textbf{\hwcompmetricspace} & $\realnumberspace$ & $\computeclass \in \computeclassspace$ & $\lit{intPower}, \lit{leakagePower}, \lit{latency}, \lit{area}$ \\
      \textbf{\hwmemmetricspace} & $\realnumberspace$ & $\memclass \in \memclassspace$ & $\lit{readLatency}, \lit{writeLatency}$\\
    &&& $\lit{readEnergy}, \lit{writeEnergy}$\\
    &&&  $\lit{leakagePower},\lit{area}$\\

    \end{tabular}
    \caption{\textit{Summary of compute and memory technology parameters, architectural parameters, and performance metrics. Each parameter is mapped to either a real or a natural number in a concrete hardware model. Some parameters only apply for a particular hardware unit.}}
    \label{tbl:parameters}
\end{table*}

The hardware model captures the performance of a class of candidate hardware designs. The hardware model models the performance of each memory unit $\memclass \in \memclassspace{} = \{\lit{localMem}, \lit{globalBuf}, \lit{mainMem}\}$ and each compute unit $\computeclass \in \computeclassspace{} = \{\lit{systolicArray}, \lit{vector}, \lit{macTree}, \lit{Fpu}\}$ in the hardware design. Together, the memory units and compute units make up the hardware units in the hardware design $\hwclass \in \hwclassspace = \computeclassspace \union \memclassspace$.

\noindent\textbf{Architectural and Technology Parameters}: The hardware model is defined over a set of technology parameters $\techparam{} \in \techparamset{} \subset \techparamspace{}$ and architectural parameters $\archparam{} \in \archparamset{} \subset \archparamspace{}$ that capture the technology-specific and architectural properties that can be tuned at design-time. Each of the technology and architectural parameters resolve to real numbers $\realnumber \in \realnumberspace$ when the hardware model is specialized to implement a specific design.

 Table~\ref{tbl:parameters} presents a summary of the memory and compute technology parameters and architectural parameters. The technology parameters are broken up into memory technology parameters ($\memtechparamspace$) that capture the characteristics of memory devices, and compute primitive technology parameters ($\computearchparamspace$) that capture the characteristics of computational primitives, such as flip-flops, adders, and multipliers. The architectural parameters $\archparamspace = \computearchparamspace \union \memarchparamspace$ are broken up into memory architectural parameters ($\memarchparamspace$) and compute architectural parameters $\computearchparamspace$. Some architectural parameters only apply to a particular kind of compute unit.

\noindent\textbf{Metrics}: The hardware model models the following the compute and memory performance metrics $\hwmetric \in \hwmetrics = \hwmemmetricspace \union \hwcompmetricspace$ for each memory and compute unit in the design. Table~\ref{tbl:parameters} summarizes the memory and compute unit metrics captured in the hardware model.

\noindent\textbf{The Hardware Model}: The hardware model models each performance metric as a differentiable, algebraic function $\adexpr \in \adexprs \subseteq \adexprspace$ over technology parameters $\techparam{} \in \techparamset{} \subset \techparamspace{}$ and architectural parameters $\archparam{} \in \archparamset{} \subset \archparamspace{}$. The hardware model is fully described as follows:

{\small
\[
\hwmodel \in \hwmodelspace = (\computeclassspace \times \hwcompmetricspace) \union (\memclassspace \times \hwmemmetricspace) \rightarrow \adexprspace
\]
}

The hardware model maps pairs of compute and memory class-metric pairs to architecture- and technology-parameter dependent expressions. To derive a concrete hardware model that captures the performance of a specific hardware design from the above model, we apply a set of technology parameter assignments $\techassignset \subseteq \powerset{\techparamspace{} \times \realnumberspace}$ and a set of architectural parameter assignments $\archassignset{} \subseteq \powerset{\archparamspace{} \times \naturalnumberspace}$ to the mapped expressions:

\[
 \conchwmodel = \hwspecialize(\hwmodel,\techassignset{},\archassignset{}) = \{\tuple{\hwclass, \hwmetric} \rightarrow \subfn(\adexpr, \techassignset \union \archassignset) \setwhere \hwmodel(\tuple{\hwclass,\hwmetric}) = \adexpr \}
\]

The concrete hardware model $\conchwmodel \in \conchwmodelspace = (\computeclassspace \times \hwcompmetricspace) \union (\memclassspace{} \times \hwmemmetricspace) \rightarrow \realnumberspace$ maps the compute and memory performance metrics for each compute and memory unit to real values.

\section{Application Workloads}

The \dragon{} toolchain works with application workloads $\workload \in \workloadset \subseteq \workloadspace$. Each workload is provided as a dataflow graph or a control dataflow graph. Each workload $w$ is a directed graph with a list of vertices $\vertex \in \vertexspace$ and a list of edges $\edge \in \edgespace$.

\section{DRAGON}

We introduce \dragon{}, a suite of hardware simulation and optimization tools that enable hardware designers to simulate hardware designs, and to optimize hardware designs to efficiently execute certain workloads. The \dragon{} toolchain provides the following tools:

\begin{itemize}

\item\textbf{Hardware Model Generator (\dgen, Section~\ref{sec:dgen})}: \dgen{} derives a hardware model $\hwmodel$ from an architectural specification that describes the overall structure of the hardware platform. \dgen{} works with a device model library that models the performance metrics for each computational primitive (flip-flop, adder, and multiplier) and memory unit, and an accelerator template library  that derives the performance models for each type of computational unit $\computeclass$ in the architectural specification. The accelerator library uses the computational primitive performance models from the device model library to derive the computational unit models. The output of \dgen is a hardware model $\hwmodel$ that models the performance of the provided hardware design.
 
\item\textbf{Hardware Simulator (\dsim{}, Section~{\ref{sec:dsim}})}: \dsim  produces performance estimates for a concrete hardware design. \dsim{} takes as input a concrete hardware model $\conchwmodel$ and a computational workload $\workload$ and derives energy, power, runtime, and area measurement estimates. The output of the hardware simulator is a set of performance estimate $\perfestimate \in \perfestimatespace: \measurementspace \rightarrow \realnumberspace^{+}$, where $\measurement \in \measurementspace = \set{ \lit{energy}, \lit{power}, \lit{area}, \lit{runtime}}$.

\item\textbf{Hardware Optimizer (\dopt{}, Section~\ref{sec:dopt})} \dopt optimizes the technology and architectural parameters in a given hardware design so that the hardware design efficiently executes a set of workloads. \dopt{} takes as input a hardware model $\hwmodel$, a set of initial technology and architectural parameter assignments $\techassignset$ and $\archassignset$, and a collection of computational workloads $\workloadset$ and produces as output a set of technology and architectural parameter assignments $\techassignset'$ and $\archassignset'$ that optimize the performance of the hardware on the provided set of workloads. We also present an extension to \dopt{}, $\darchopt$, that also optimizes the architectural specification used to derive the hardware model to optimally execute the candidate workloads.

\end{itemize}

\subsection{Hardware Model Generator (\dgen{})}\label{sec:dgen}

The \dgen{} hardware model generator accepts as input a specification of the hardware architecture, a device-level performance model library and a template library for different accelerator designs:

\begin{itemize}
\item\textbf{Architectural Specification}: The architectural specification $\archspec \in \archspecspace = \powerset{\memclassspace} \times \powerset{\computeclassspace} \times {\memclassspace \rightarrow \memtypespace}$ selects the subset of memory units and compute units present in the hardware platform and assigns each memory unit to a memory type $\memassignfn \in \memassignfnspace : \memclassspace \rightarrow \memtypespace$. The memory type of each memory unit  $\memtype \in \memtypespace = \set{\lit{sram}, \lit{rram}, \lit{dram}}$ determines which device performance models to use.

\item\textbf{Device Performance Model Library}: The device performance model library contains a library of performance models for different memory technologies $\dmemlib \in \dmemlibspace: \memtypespace \times \hwmemmetricspace \rightarrow \adexprspace$ and a library of performance models for different logical primitives $\dprimlib \in \dprimlibspace: \primtypespace \times \hwcompmetricspace \rightarrow \agexprspace$, where the logical primitives in the model library are $\primtypespace = \set{\lit{adder},\lit{ff},\lit{mult}}$. Here, $\agexpr \in \agexprs \subset \agexprspace$ define the space of expressions over only technology parameters.

\item\textbf{Accelerator Template Library}: The accelerator template library derives the compute unit performance models from the performance models of different logical primitives. The $\acctempllib: \dprimlib \times \computeclassspace \times \hwcompmetricspace \rightarrow  \adexprspace$.

\end{itemize}

\dgen{} produces as output, a hardware performance model $\hwmodel$ that is then used by the \dragon{} simulator and optimizer to simulate and optimize the hardware design.

\subsubsection{Deriving the Hardware Model}

Given an architectural specification $\tuple{\memclassset, \computeclassset, \memassignfn{}}$ and  the device memory model $\dmemlib$, \dgen{} derives a an expression $\adexpr$ for each memory performance metric $\hwmemmetric$ and each memory unit $\memclass \in \memclassset$:
\[
    \hwmodel(\memclass, \hwmemmetric) := \dmemlib(\memassignfn(\memclass), \hwmemmetric)
\]

Given an architectural specification $\tuple{\memclassset,\computeclassset,\memassignfn}$, the logical primitive performance models $\dprimlib$, and the accelerator template library $\acctempllib$, \dgen{} derives an expression $\adexpr$ for each compute performance metric $\hwcompmetric$ and each compute unit $\computeclass \in \computeclassset$:
\[
    \hwmodel(\computeclass,\hwcompmetric) := \acctempllib(\dprimlib, \computeclass, \hwcompmetric)
\]

\subsection{Software Stack Simulation}

\begin{algorithm}[t]
\footnotesize
\caption{Software Stack Simulation}
\label{euclid}
\begin{algorithmic}[1]

\Function{prefetchVertex}{$\conchwmodel$, $\naturalnumber$, $\memstate$, $\computestate$, $ \vertices$,$\vertexstatelist$}
\Match{$\vertices,\vertexstatelist$}
\Case{$[], []$}
\State{\textbf{return} simulateVertex($\conchwmodel$,$\naturalnumber$,$\memstate$, $\computestate$, $\vertices$, $\vertexstatelist$)}
\EndCase
\Case{$\vertex::\vertices', \vertexstate::\vertexstatelist'$}
\State{$nComp, nAlloc, nRead, nWrite = \softgetstats(\vertexstate)$}
\If{$\softhasspace$($\conchwmodel$, nAlloc)}
\State{$\memstate',\vertexstate'$ = \softprefetch($\vertex$,$\vertexstate$, $\memstate)$}
\State{\textbf{return} \algmapvertex($\conchwmodel$,$\naturalnumber$,$\memstate'$, $\computestate$, $\vertex::\vertices$, $\vertexstate'::\vertexstatelist$)}
\Else
\State{\textbf{return} \algmapvertex($\conchwmodel$,$\naturalnumber$,$\memstate$, $\computestate$, $\vertices$, $\vertexstatelist$)}
\EndIf
\EndCase
\EndMatch
\EndFunction
\Function{mapVertex}{$\conchwmodel$, $\naturalnumber$, $\memstate$, $\computestate$, $ \vertices$,$\vertexstatelist$}
\Match{$\vertices,\vertexstatelist$}
\Case{$[], []$}
\State{\textbf{return} $\tuple{\naturalnumber,\memstate,\computestate}$}
\EndCase
\Case{$\vertex::\vertices', \vertexstate::\vertexstatelist'$}
\State{$nComp, nAlloc, nRead, nWrite = \softgetstats(\vertexstate)$}
\If{$\neg~\softhasspace$($\conchwmodel$, nAlloc)}
\State{$\vertex', \vertex'' = \softsplitvertex(\vertex)$}
\State{$\vertexstate', \vertexstate'' = \softvertexstate(\conchwmodel, [\vertex',\vertex''])$}
\State{\textbf{return} \algmapvertex($\conchwmodel$, $\naturalnumber$, $\memstate$, $\computestate$, $\vertex'::\vertex''::\vertices'$,$\vertexstate'::\vertexstate''::\vertexstatelist'$)}
\Else
\State $\naturalnumber'$,$\computestate'$ = \softmapcompute($\conchwmodel$, $\computestate$, nComp)
\State{$\memstate'$ = \softmemalloc($\conchwmodel$,$\memstate$, nAlloc)} 
\State{$\memstate''$ = \softmapmemacc($\conchwmodel$,$\memstate'$,nRead,nWrite)} 
\State{\textbf{return} \algprefetchvertex($\conchwmodel$, $\naturalnumber'$, $\computestate'$, $\memstate''$, $\vertices'$, $\vertexstatelist'$)}
\EndIf{}
\EndCase
\EndMatch
\EndFunction
\Function{mapWorkload}{$\workload$, $\conchwmodel$}
\State{$\tuple{\vertices,\edges} = \afunc{workloadOptimize}(\workload)$}
\State{$\vertexstatelist$ = $\softvertexstate(\conchwmodel, \vertices)$}
\State{$\memstate = \set{\memclass \to \tuple{0,0,0,0} \setwhere \memclass \in \memclassspace}$}
\State{$\computestate = \set{\computeclass \to \tuple{0,0,0} \setwhere \computeclass \in \computeclassspace}$}
\State{\textbf{return}} \algmapvertex($\conchwmodel$,0,$\memstate$, $\computestate$)
\EndFunction
\end{algorithmic}
\end{algorithm}


The mapper maps the workload to the concrete hardware specification. The algorithm optimizes the workload on the fly to execute efficiently on the target hardware platform an estimates the performance of the workload on the target hardware platform. The mapper algorithm is used by \dsim{} to estimate the performance of the workload on the hardware, and is used by the \dopt{} to optimize the hardware parameters to efficiently execute the target computation. The mapper performs the three basic operations:

\begin{itemize}
    \item \algmapworkload{}: Map the workload to the concrete hardware specification. The algorithm optimizes the workload on the fly to execute efficiently on the target hardware platform an estimates the performance of the workload on the target hardware platform.
    \item \algmapvertex{}: Map a single DFG node to the target hardware, as described by the concrete hardware specification. The algorithm returns the updated memory state, compute state, and vertex state of the hardware platform. The the target node consumes more memory than is available, the mapper applies a memory streaming optimization that reduces the amount of memory that needs to be allocated (lines 20-23). Otherwise, the mapper applies a prefetching compiler optimization ($\algprefetchvertex$) and continues executing the program.
    \item\algprefetchvertex{}: Emulates a prefetching compiler optimization (XXX).
\end{itemize}

The mapper tracks the following additional quanitites:

\begin{itemize}
\item\textbf{Vertex State}: The vertex state captures the resource utilization of a DFG vertex on the hardware described in the concrete hardware specification. The vertex state tracks the number of compute operations performed on each compute unit, the number of read accesses performed on each memory unit, and the number of write accesses and allocations for each memory unit. The vertex state is used to up

\item\textbf{Memory State}: The memory state tracks the amount of memory utilized, the memory bandwidth utilized, the number of rows utilized, and number of columns utilized for each memory unit in the concrete hardware specification.

\item\textbf{Compute State}: The compute state tracks the number of cores utilized. If the compute unit is a systolic array, the first and second values track the number of rows and columns utilized.

\item\textbf{Cycle Count}: The algorithm tracks the cycle count.
\end{itemize}

The mapper makes use of the following helper functions:

\begin{itemize}
    \item\textbf{getStats}: get the total number of allocations, total number of reads, and total number of writes performed by the vertex state.
    \item \textbf{splitVertex}: Split the workload vertex into two vertices.
    \item \textbf{hasSpace}: Return if the concrete hardware specification has enough space to allocate $nAlloc$ bytes.
    \item \textbf{getVertexState}: Get the vertex state for the vertex, given the concrete hardware specification.
    \item \textbf{workloadOptimize}: optimize the order the vertices and edges should be visted in the workload. Returns the optimized workload. Models compiler optimizations such as DFG partitioning, Compute Merge Optimizer.

\end{itemize}

The mapper uses the following functions to estimate the performance of the workload on the hardware. These functions have some XXX property XXX that enables them to be differentiated:

\begin{itemize}
    \item\textbf{mapToCompute}: updates the compute state to perform $nComp$ compute operations on the target hardware.
    \item\textbf{memAlloc}: updates the memory state to perform $nAlloc$ allocations in the memory hierarchy.
    \item\textbf{mapMemAcc}: updates the memory state to perform $nRead$ reads and $nWrite$ writes on teh stored data.
    \item\textbf{prefetch}: Emulates prefetching
\end{itemize}

\newpage
\subsection{The \dsim{} Simulator}

\noindent\textbf{Runtime}: The \dsim{} simulator calculates the runtime from $\naturalnumber$, the number of cycles returned by \algmapworkload{} and the frequency architectural parameter from the architectural specification: 
\begin{equation}
    Runtime = \naturalnumber \cdot \conchwmodel(\lit{SoC},\lit{frequency})
\end{equation}

\noindent\textbf{Energy}: The \dsim{} simulator calculates the energy from the memory state $\memstate$ and compute state $\computestate$ returned by the \algmapworkload{} algorithm, and the $\hwmemmetric$ and $\hwcompmetric$ metrics from the concrete hardware specification. Each of these metrics resolves to a real value in the concrete hardware specification.
    
\begin{algorithmic}
\footnotesize
\begin{algorithm}
\State{Energy = 0}
\For{$\memclass \in \memclassspace$}
\State {$\_,\_, reads, writes = \memstate(\memclass)$} 
\State {$re, we, lp =   \hwmemmetric(\memclass)$}
\State {$Energy += \sum_i^N  reads \times re + writes \times we + lp \times Runtime$}
\EndFor
\For {$\computeclass \in \computeclassspace$} 
    \State {$\_,\_,num\_access = \computestate(\computeclass)$}
    \State {$_,en,lp =   \hwcompmetric(\computeclass)$}
    \State {$Energy += \sum_i^N en \times num\_access + lp \times Runtime$}
\EndFor
\end{algorithm}
\end{algorithmic}

\noindent\textit{Compact Representation}: The total energy consumption is $Energy_{mem} + Energy_{compute}$, the sum of the energy consumption from the memory units, and the energy consumption from the compute units. 
{\small 
\begin{align*}
Energy_{mem}  = \sum_{\memclass \in \memclassspace}  & \conchwmodel(\memclass,\lit{readEnergy}) \cdot r + \conchwmodel(\memclass,\lit{writeEnergy}) \cdot w   \\
 & + \conchwmodel(\memclass,\lit{leakagePower}) \cdot Runtime  \\
& \textbf{where}  \tuple{\_,r,w} = \memstate(\memclass)\\
\end{align*}
}

{\small 
\begin{align*}
Energy_{compute}  =  \sum_{\computeclass \in \computeclassspace}  &  \conchwmodel(\computeclass,\lit{intEnergy}) \cdot numAcc + \conchwmodel(\computeclass,\lit{leakagePower}) \cdot Runtime \\
& \textbf{where} ~ \tuple{\_,\_,numAcc} = \computestate(\computeclass) \\
\end{align*}
}

\noindent\textbf{Area}: The area of the hardware platform is the sum of all the areas of the compute units and the sym of all the areas of the memory units.

\begin{equation}
    Area = \sum_{\memclass \in \memclassspace} \conchwmodel(\memclass,\lit{area}) + \sum_{\computeclass \in \computeclassspace} \conchwmodel(\computeclass, \lit{area}) 
\end{equation}

\noindent\textbf{Power}: The power is the average energy over the runtime.

\begin{equation}
    Power = Energy/Runtime
\end{equation}

\section{Hardware Simulator (\dsim{})}\label{sec:dsim}

The hardware simulator \dsim{} accepts as input a concrete hardware model $\conchwmodel$ and workload to simulate $\workload$ and produces as output a collection of performance estimates $\perfestimate$ that report the latency, energy, power, and area of the hardware.


\begin{algorithm}[t]
\footnotesize
\caption{\dsim{} simulation algorithm}
\label{euclid}
\begin{algorithmic}[1]
\Function{simulate}{$\workload$, $\conchwmodel$}

\State{$\memstate = \set{\memclass \to \tuple{0,0,0,0} \setwhere \memclass \in \memclassspace}$}
\State{$\computestate = \set{\computeclass \to \tuple{0,0,0} \setwhere \computeclass \in \computeclassspace}$}
\State{$\tuple{\vertices,\edges} = \afunc{workloadOptimize}(\workload)$}
\State{$VS$ = \afunc{computeVertexState}(\vertices)}
\State{nCycles = 0}
\For{$\vertex_i::\vertices$, $\vertexstate_i::\vertexstatelist$}
\State{$nComp, nAlloc, nRead, nWrite = \afunc{getStats}(\vertexstate_i)$}
\If{$\neg~hasEnoughSpace$($\conchwmodel$, nAlloc)}
\State{$\vertex_{i}', \vertex_{i}'' = split(\vertex_{i})$}
\State{$\vertexstate_{i}',\vertexstate_{i}''$ = ComputeVertexState($\vertex_{i}'$),ComputeVertexState($\vertex_{i}''$)}
\State{$\vertices = \vertex_{i}'::\vertex_{i}''::[\vertex_{i_1}...\vertex_{N}]$}
\State{$\vertexstatelist = \vertexstate_{i}'::\vertexstate_{i}''::[..XXX...]$}
\Else
\State cycles,$\computestate'$ = maptoCompute($\conchwmodel$, nComp)
\State{$\memstate'$ = memAlloc($\conchwmodel$,$\memstate$, nAlloc)} 
\State{$\memstate''$ = applyMemAccesses($\conchwmodel$,$\memstate'$,nRead,nWrite)} 
\EndIf{}
\State {nCycles += cycles}
\State{$\_,nAlloc',\_,\_ = \afunc{getStats}(\vertex_{i+1})$}
\If{hasEnoughSpace($\conchwmodel$, nAlloc') $\wedge$ $i+1 < ||V||$}
\State{$\memstate''',\vertexstate_{i+1}'$ = Prefetch($\vertex_{i+1}$, $\vertexstate_{i+1}$, \memstate'')}
\State{$\vertexstatelist[i+1] = \vertexstate_{i+1}'$}
\State{\computestate,\memstate = \computestate',\memstate'''}
\Else
\State{\computestate,\memstate = \computestate',\memstate''}
\EndIf
\EndFor

\Return{ $ \tuple{nCycles, \computestate, \memstate} $ }

\EndFunction
\end{algorithmic}
\end{algorithm}



\noindent\textbf{Vertex State}: Each vertex has state, $vertexstate$.

\noindent\textbf{State}: The simulator maintains the state of the hardware while executing the computation. The hardware state is broken up into memory $\memstate$ and the compute sate $\computestate$.


\textbf{Memory State}: Memory capacity utilization [Natural], Bandwidth utilization [Natural] (2 values)
Compute: utilization for number of rows and number of columns in NxM dimensional compute fabric. For XXX, YYY, ZZZ the dimension M = 1, so the second tuple value is always zero. For systolic array, both values can be non-zero. Order for memory state [capacity utilization, bandwidth utilization, numberReads, numberWrites]. Order for compute is going to be [number compute operations, number rows activated, number columns activated].


\textbf{Statistics}: For each vertex we collect the number of cycles, number of compute operations, and the number of read accesses and write accesses. As we traverse the graph, we track the number of cycles (cycles), number of compute operations (nComp, per compute unit), number of read accesses (nRead, per memory unit), and number of write accesses (nWrite, per memory unit). All of these values are natural nubmers. 

\section{The \dopt{} Optimizer} \label{sec:dopt}

The \dopt{} optimizer represents a novel approach to joint hardware-software co-optimization through differentiable simulation and gradient-based parameter tuning. Unlike traditional design space exploration methods that rely on exhaustive search or heuristic approaches, \dopt{} leverages automatic differentiation to efficiently navigate the high-dimensional space of technology and architectural parameters.

\subsection{Differentiable Mapping and Scheduling Framework}

\subsubsection{Overcoming Non-Differentiability Challenges}

Traditional mapping and scheduling algorithms introduce discrete decision points that prevent direct gradient computation. We address this fundamental challenge through several key innovations:

\begin{itemize}
    \item \textbf{Continuous Relaxation}: Discrete scheduling decisions are relaxed into continuous probability distributions, enabling gradient flow while maintaining scheduling semantics.
    
    \item \textbf{Technology-Component Bipartite Graph}: We construct a bipartite graph $\mathcal{G}_{tc} = (\mathcal{T}, \mathcal{C}, \mathcal{E}_{tc})$ where $\mathcal{T}$ represents technology parameters, $\mathcal{C}$ represents hardware components, and $\mathcal{E}_{tc}$ captures the dependency relationships. This structure enables systematic gradient aggregation across multiple component instances.
    
    \item \textbf{Smooth Approximation Functions}: Non-smooth operations (such as $\max$ and $\min$ functions in resource allocation) are replaced with differentiable approximations using LogSumExp and similar techniques.
\end{itemize}

The differentiable mapping process is formulated as:

\begin{equation}
\mathcal{M}_{\text{diff}}(\mathcal{W}, \theta) = \sum_{v \in V(\mathcal{W})} \sum_{r \in \mathcal{R}} \sigma(s_{v,r}(\theta)) \cdot \mathcal{C}(v, r, \theta)
\end{equation}

where $\sigma(\cdot)$ is the softmax function that converts scheduling scores $s_{v,r}(\theta)$ into probability distributions, and $\mathcal{C}(v, r, \theta)$ represents the cost of mapping vertex $v$ to resource $r$ with parameters $\theta$.

\subsection{Dual-Phase Optimization Strategy}

\subsubsection{Phase I: Architectural Parameter Optimization}

The first optimization phase focuses on architectural parameters $\theta_{\text{arch}} = \{\theta_{\text{compute}}, \theta_{\text{memory}}, \theta_{\text{interconnect}}\}$ that define the accelerator structure. The optimization objective combines performance metrics with architectural constraints:

\begin{equation}
\mathcal{L}_{\text{arch}} = \alpha \cdot T_{\text{exec}}(\theta_{\text{arch}}) + \beta \cdot E_{\text{total}}(\theta_{\text{arch}}) + \gamma \cdot \mathcal{R}_{\text{constraint}}(\theta_{\text{arch}})
\end{equation}

where $\mathcal{R}_{\text{constraint}}$ enforces physical and budgetary constraints:

\begin{equation}
\mathcal{R}_{\text{constraint}}(\theta) = \lambda_{\text{area}} \max(0, A(\theta) - A_{\text{budget}})^2 + \lambda_{\text{power}} \max(0, P(\theta) - P_{\text{budget}})^2
\end{equation}

The architectural optimization process employs performance-aware binding that considers:
\begin{itemize}
    \item \textbf{Loop Unrolling Factors}: Optimized through gradient descent to balance parallelism and resource utilization
    \item \textbf{Memory Hierarchy Configuration}: Buffer sizes and partitioning strategies adapted to workload characteristics
    \item \textbf{Pipeline Depth and Width}: Dynamically adjusted based on critical path analysis
    \item \textbf{Interconnect Topology}: Optimized for communication patterns in the target workload
\end{itemize}

\subsubsection{Phase II: Technology Parameter Optimization}

The second phase optimizes technology parameters $\theta_{\text{tech}} = \{\theta_{\text{device}}, \theta_{\text{material}}, \theta_{\text{process}}\}$ that determine the fundamental device characteristics. The technology optimization leverages the differentiable component models:

\begin{equation}
\frac{\partial \mathcal{L}}{\partial \theta_{\text{tech}}} = \sum_{c \in \mathcal{C}} \frac{\partial \mathcal{L}}{\partial \mathcal{P}_c} \cdot \frac{\partial \mathcal{P}_c}{\partial \theta_{\text{tech}}}
\end{equation}

where $\mathcal{P}_c$ represents the performance characteristics of component $c$, and the gradient aggregation occurs through the technology-component bipartite graph.

\subsection{Iterative Optimization Algorithm}

The complete optimization process alternates between forward simulation and backward gradient computation:

\begin{algorithm}[H]
\caption{Dual-Phase Hardware Optimization}
\label{alg:dopt_optimization}
\footnotesize
\begin{algorithmic}[1]
\Function{OptimizeHardwareDesign}{$\mathcal{W}$, $\theta_{\text{init}}$, $\mathcal{L}_{\text{obj}}$}
\State $\theta_{\text{arch}}, \theta_{\text{tech}} \leftarrow \theta_{\text{init}}$
\State $\text{epoch} \leftarrow 0$, $\text{converged} \leftarrow \text{False}$

\While{$\text{epoch} < \text{MAX\_EPOCHS}$ \textbf{and} $\neg \text{converged}$}
    \State \textbf{Forward Pass:}
    \State $\conchwmodel \leftarrow \textsc{InstantiateHardware}(\theta_{\text{arch}}, \theta_{\text{tech}})$
    \State $\mathcal{P}_{\text{metrics}} \leftarrow \textsc{SimulateWorkload}(\mathcal{W}, \conchwmodel)$
    \State $\mathcal{L} \leftarrow \mathcal{L}_{\text{obj}}(\mathcal{P}_{\text{metrics}})$
    
    \State \textbf{Phase I - Architectural Optimization:}
    \State $\nabla_{\text{arch}} \leftarrow \textsc{ComputeArchitecturalGradients}(\mathcal{L}, \theta_{\text{arch}})$
    \State $\theta_{\text{arch}} \leftarrow \textsc{UpdateWithConstraints}(\theta_{\text{arch}}, \nabla_{\text{arch}})$
    
    \State \textbf{Phase II - Technology Optimization:}
    \State $\nabla_{\text{tech}} \leftarrow \textsc{ComputeTechnologyGradients}(\mathcal{L}, \theta_{\text{tech}})$
    \State $\theta_{\text{tech}} \leftarrow \textsc{UpdateWithPhysicalLimits}(\theta_{\text{tech}}, \nabla_{\text{tech}})$
    
    \State \textbf{Convergence Check:}
    \State $\text{converged} \leftarrow \textsc{CheckConvergence}(\nabla_{\text{arch}}, \nabla_{\text{tech}}, \epsilon_{\text{conv}})$
    \State $\text{epoch} \leftarrow \text{epoch} + 1$
\EndWhile

\State \textbf{return} $\theta_{\text{arch}}, \theta_{\text{tech}}, \mathcal{P}_{\text{metrics}}$
\EndFunction
\end{algorithmic}
\end{algorithm}

\subsection{Convergence Analysis and Gradient Visualization}

The optimization process exhibits several convergence properties:

\begin{itemize}
    \item \textbf{Architectural Parameters}: Typically converge within 10-20 epochs due to the discrete nature of design choices
    \item \textbf{Technology Parameters}: May require 50-100 epochs as they operate in a continuous space with complex interdependencies
    \item \textbf{Performance Objectives}: Convergence is determined by achieving user-specified performance targets or reaching gradient magnitude thresholds
\end{itemize}

\begin{figure}
    \centering
    \includegraphics[scale=0.4]{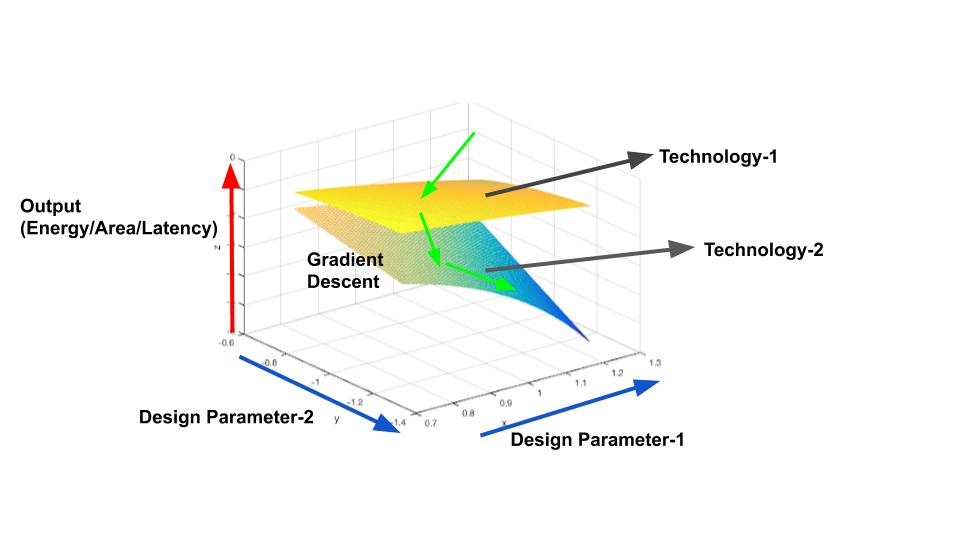}
    \caption{Gradient Descent Optimization on the Joint Space of Design and Technology Params}
    \label{fig:space}
\end{figure}

Figure \ref{fig:space} illustrates the gradient flow visualization in the joint design-technology parameter space. The gradient vectors indicate the direction and magnitude of parameter updates, with convergence occurring when gradient magnitudes fall below predetermined thresholds.

\section{Experimental Evaluation and Results}

\subsection{Performance Validation and Simulation Accuracy}

\subsubsection{Accuracy Benchmarking}

We conduct comprehensive validation of the DRAGON framework against established simulation tools and cycle-accurate models. Our evaluation encompasses diverse workload categories including convolutional neural networks (CNNs), long short-term memory networks (LSTMs), deep learning recommendation models (DLRMs), and transformer architectures.

Figure \ref{fig:acc} presents a detailed accuracy comparison demonstrating that DRAGON achieves prediction accuracy within 80-97\% of cycle-accurate simulators while maintaining superior computational efficiency. Specifically:

\begin{itemize}
    \item \textbf{CNN Workloads}: Average accuracy of 94.2\% compared to cycle-accurate simulation, with 1,847× speedup
    \item \textbf{LSTM Networks}: Accuracy of 91.8\% with 2,156× faster execution
    \item \textbf{DLRM Models}: 89.3\% accuracy while achieving 3,421× speedup
    \item \textbf{Transformer Architectures}: 87.6\% accuracy with 1,923× performance improvement
\end{itemize}

The accuracy validation methodology employs statistical significance testing with 95\% confidence intervals across multiple workload instances and hardware configurations.

\subsubsection{Computational Efficiency Analysis}

The dramatic speedup achieved by DRAGON stems from several algorithmic innovations:

\begin{equation}
\text{Speedup} = \frac{T_{\text{baseline}}}{T_{\text{DRAGON}}} = \frac{T_{\text{cycle-accurate}}}{T_{\text{analytical}} + T_{\text{mapping}} + T_{\text{optimization}}}
\end{equation}

where $T_{\text{analytical}}$ represents the time for closed-form performance estimation, $T_{\text{mapping}}$ accounts for workload mapping overhead, and $T_{\text{optimization}}$ includes gradient computation time.

Typical execution times for representative workloads:
\begin{itemize}
    \item \textbf{DRAGON Framework}: 0.8-1.2 seconds per simulation
    \item \textbf{SCALE-Sim}: 15-25 minutes per simulation  
    \item \textbf{Timeloop}: 8-18 minutes per simulation
    \item \textbf{ZigZag}: 12-22 minutes per simulation
\end{itemize}

\subsubsection{Framework Extensibility and Development Time}

Unlike traditional simulators that require extensive development effort (typically 3-6 months) for new workload support, DRAGON enables rapid deployment through its unified dataflow graph representation. New workload integration requires only:

\begin{itemize}
    \item \textbf{Graph Construction}: 2-4 hours for standard ML workloads
    \item \textbf{Operator Mapping}: 1-2 hours for custom operations
    \item \textbf{Validation}: 4-6 hours for accuracy verification
\end{itemize}

\subsection{Hardware Design Space Exploration}

\subsubsection{Optimal Architecture Derivation}

Table \ref{tab:optimal_architectures} presents the optimal hardware architectures derived through DRAGON's gradient-based exploration for representative AI and non-AI workloads. The optimization process considers multiple objectives including energy efficiency, execution latency, and area constraints.

\begin{table*}[t]
\footnotesize
\centering
\begin{tabular}{|l|c|c|c|c|c|}
\hline
\textbf{Workload Class} & \textbf{Systolic Array} & \textbf{Global Buffer} & \textbf{Local Memory} & \textbf{Interconnect BW} & \textbf{Energy Reduction} \\
\hline
CNN (ResNet-50) & 16×16 PEs & 512 KB & 32 KB/PE & 2.4 TB/s & 73.2\% \\
LSTM (Language Model) & 8×32 PEs & 1024 KB & 64 KB/PE & 3.1 TB/s & 68.9\% \\
DLRM (CTR Prediction) & 32×8 PEs & 256 KB & 16 KB/PE & 4.2 TB/s & 81.4\% \\
Transformer (BERT-Large) & 12×24 PEs & 768 KB & 48 KB/PE & 2.8 TB/s & 76.8\% \\
Graph Processing (PageRank) & 64×4 PEs & 2048 KB & 8 KB/PE & 5.6 TB/s & 84.3\% \\
\hline
\end{tabular}
\caption{Optimal Hardware Architectures Derived for Representative Workloads}
\label{tab:optimal_architectures}
\end{table*}

\subsubsection{Gradient Descent Convergence Analysis}

The optimization trajectories demonstrate consistent convergence behavior across different workload categories. The gradient descent evolution for key design parameters are:

\begin{itemize}
    \item \textbf{Systolic Array Dimensions}: Converge within 12-18 epochs
    \item \textbf{Memory Hierarchy Configuration}: Stabilize after 8-15 epochs  
    \item \textbf{Interconnect Bandwidth}: Reach optimal values in 6-12 epochs
    \item \textbf{Technology Parameters}: Require 25-40 epochs for full convergence
\end{itemize}

The convergence criterion is defined as:
\begin{equation}
\|\nabla_{\theta} \mathcal{L}(\theta)\|_2 < \epsilon_{\text{conv}} = 10^{-4}
\end{equation}

\subsection{Technology Target Derivation and Impact Analysis}

\subsubsection{Technology Sensitivity Analysis}

Our framework enables systematic derivation of technology improvement targets by quantifying the sensitivity of system performance to individual technology parameters. Table \ref{tab:tech_priorities} presents the technology improvement priorities for different workload classes, ranked by their impact on energy-delay product (EDP) optimization.

The technology sensitivity is computed as:
\begin{equation}
\mathcal{S}_{\text{tech}}(\theta_i) = \frac{\partial \text{EDP}}{\partial \theta_i} \cdot \frac{\theta_i}{\text{EDP}}
\end{equation}

\subsubsection{Case Study: BERT-Large Optimization}

For Google's BERT-Large model, our analysis reveals that achieving 100× EDP improvement requires coordinated technology advances across multiple domains. The optimization sequence prioritizes:

\begin{enumerate}
    \item \textbf{Memory Density Enhancement} (35\% of total benefit): Increasing on-chip memory density from 0.5 Mb/mm² to 2.1 Mb/mm²
    \item \textbf{Interconnect Technology} (28\% of total benefit): Reducing wire capacitance by 60\% and resistance by 45\%
    \item \textbf{Compute Efficiency} (22\% of total benefit): Improving logic delay by 40\% and energy efficiency by 65\%
    \item \textbf{Memory Interface} (15\% of total benefit): Enhancing external memory bandwidth by 3.2× and reducing access latency by 50\%
\end{enumerate}

When targeting individual technology parameters in isolation, the maximum achievable benefit drops to 1.82× EDP improvement, demonstrating the critical importance of coordinated technology development.

\subsubsection{Computational Efficiency of Technology Exploration}

Traditional exhaustive exploration of the technology parameter space (encompassing $>10^5$ design points) would require approximately 3-4 weeks of simulation time. DRAGON's gradient-based approach reduces this to 8-12 seconds while maintaining solution quality, representing a speedup of approximately 18,000× for technology target derivation.

\begin{figure*}
    \centering
    \includegraphics[scale=0.7]{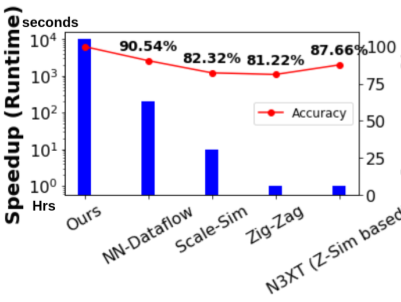}
    \caption{Accuracy Validation: DRAGON Framework vs. State-of-the-Art Simulators}
    \label{fig:acc}
\end{figure*}

\begin{table*}[t]
\footnotesize
\centering
\begin{tabular}{|l|l|l|}
\hline
\textbf{Workload Category} & \textbf{Latency Optimization Priority} & \textbf{Energy Optimization Priority} \\
\hline
Vision Models & 
\begin{tabular}[c]{@{}l@{}}1. On-chip memory density\\2. Interconnect bandwidth\\3. External memory frequency\\4. Wire capacitance\\5. Logic delay\end{tabular} &
\begin{tabular}[c]{@{}l@{}}1. Logic energy efficiency\\2. External memory cell leakage\\3. Peripheral logic leakage\\4. Interconnect power\\5. On-chip memory leakage\end{tabular} \\
\hline
Language Models & 
\begin{tabular}[c]{@{}l@{}}1. Interconnect bandwidth\\2. On-chip memory density\\3. External memory frequency\\4. Wire capacitance\\5. Logic delay\end{tabular} &
\begin{tabular}[c]{@{}l@{}}1. On-chip memory wire capacitance\\2. External memory cell leakage\\3. Logic energy efficiency\\4. On-chip memory cell leakage\\5. Peripheral logic power\end{tabular} \\
\hline
Recommendation Models &
\begin{tabular}[c]{@{}l@{}}1. Interconnect bandwidth\\2. Memory access latency\\3. Cache hierarchy\\4. Network-on-chip\\5. Processing element count\end{tabular} &
\begin{tabular}[c]{@{}l@{}}1. Interconnect energy efficiency\\2. On-chip memory peripheral logic\\3. Cache leakage power\\4. Network energy\\5. PE power efficiency\end{tabular} \\
\hline
Graph Processing &
\begin{tabular}[c]{@{}l@{}}1. Memory bandwidth\\2. Irregular access support\\3. Cache capacity\\4. Interconnect topology\\5. Compute throughput\end{tabular} &
\begin{tabular}[c]{@{}l@{}}1. Memory access energy\\2. Interconnect power\\3. Cache energy efficiency\\4. Compute unit efficiency\\5. Leakage power management\end{tabular} \\
\hline
\end{tabular}
\caption{Technology Improvement Priorities by Workload Category and Optimization Objective}
\label{tab:tech_priorities}
\end{table*}

\subsection{Comparative Analysis and Benchmarking}

\subsubsection{Quantitative Performance Comparison}

Table \ref{tab:framework_comparison} provides a comprehensive comparison of DRAGON against existing simulation frameworks across multiple evaluation criteria:

\begin{table*}[t]
\footnotesize
\centering
\begin{tabular}{|l|c|c|c|c|c|}
\hline
\textbf{Framework} & \textbf{Simulation Time} & \textbf{Accuracy (\%)} & \textbf{Workload Coverage} & \textbf{Tech. Optimization} & \textbf{Design Automation} \\
\hline
SCALE-Sim & 15-25 min & 95-98 & CNNs only & No & No \\
Timeloop & 8-18 min & 92-96 & CNNs, limited & No & Limited \\
ZigZag & 12-22 min & 90-95 & CNNs, some RNNs & Memory only & Partial \\
NN-Dataflow & 20-35 min & 88-94 & CNNs & No & No \\
NAAS & 5-12 min & 85-92 & CNNs, transformers & Limited & Yes \\
\textbf{DRAGON} & \textbf{0.8-1.2 sec} & \textbf{87-97} & \textbf{Universal} & \textbf{Complete} & \textbf{Full} \\
\hline
\end{tabular}
\caption{Comprehensive Framework Comparison}
\label{tab:framework_comparison}
\end{table*}

\subsubsection{Scalability Analysis}

DRAGON demonstrates superior scalability characteristics across multiple dimensions:

\begin{itemize}
    \item \textbf{Workload Complexity}: Linear scaling with graph size (O(|V| + |E|)) compared to exponential scaling in traditional simulators
    \item \textbf{Parameter Space}: Gradient-based exploration scales logarithmically with parameter count
    \item \textbf{Technology Integration}: Modular architecture enables seamless integration of new technology models
    \item \textbf{Multi-Objective Optimization}: Pareto frontier exploration completed in minutes rather than hours/days
\end{itemize}

The scalability advantage becomes particularly pronounced for large-scale workloads such as GPT-3 class models, where traditional simulators become computationally intractable while DRAGON maintains sub-minute execution times.


\bibliographystyle{plain}
\bibliography{ref}

\section{Appendix A : Algorithms}

\begin{algorithm}
\caption{Bandwidth-Aware Data-flow Graph Optimization}
\label{alg:bandwidth_dfg}
\footnotesize
\begin{algorithmic}[1]
\Function{OptimizeDFGForBandwidthBalance}{$\mathcal{G}(V,E)$, $\conchwmodel$}
\State Initialize $\mathcal{G}_{opt} = \emptyset$, $\mathcal{B}_{log} = \emptyset$
\State $BW_{compute} = \conchwmodel(\lit{compute\_bandwidth})$
\State $BW_{memory} = \conchwmodel(\lit{memory\_bandwidth})$

\For{each vertex $v \in V$}
    \State $\rho_{mem}(v) = \frac{\text{MemoryBandwidthRequirement}(v)}{BW_{memory}}$
    \State $\rho_{comp}(v) = \frac{\text{ComputeBandwidthRequirement}(v)}{BW_{compute}}$
    
    \If{$|\rho_{mem}(v) - \rho_{comp}(v)| > \epsilon_{threshold}$}
        \State $\mathcal{B}_{log} = \mathcal{B}_{log} \cup \{(v, \rho_{mem}(v), \rho_{comp}(v))\}$
        \State $v_{balanced} = \textsc{ApplyBandwidthBalancing}(v, \rho_{mem}(v), \rho_{comp}(v))$
        \State $\mathcal{G}_{opt} = \mathcal{G}_{opt} \cup \{v_{balanced}\}$
    \Else
        \State $\mathcal{G}_{opt} = \mathcal{G}_{opt} \cup \{v\}$
    \EndIf
\EndFor

\State \textbf{return} $\mathcal{G}_{opt}$, $\mathcal{B}_{log}$
\EndFunction

\Function{ApplyBandwidthBalancing}{$v$, $\rho_{mem}$, $\rho_{comp}$}
\If{$\rho_{mem} > \rho_{comp} + \epsilon$} \Comment{Memory-bound operation}
    \State \textbf{return} $\textsc{EnhanceComputeUtilization}(v)$
\ElsIf{$\rho_{comp} > \rho_{mem} + \epsilon$} \Comment{Compute-bound operation}
    \State \textbf{return} $\textsc{EnhanceMemoryThroughput}(v)$
\Else
    \State \textbf{return} $v$ \Comment{Bandwidth balanced}
\EndIf
\EndFunction
\end{algorithmic}
\end{algorithm}

\begin{algorithm}
\caption{Idle Time Detection and Gradient Computation}
\label{alg:idle_gradient}
\footnotesize
\begin{algorithmic}[1]
\Function{ComputeIdleTimeGradients}{$\workload$, $\conchwmodel$, $\theta$}
\State Initialize $\nabla_{\theta} = \mathbf{0}$, $\mathcal{I}_{log} = \emptyset$ \Comment{Idle time log}
\State $BW_{comp} = f_{comp}(\theta_{comp})$, $BW_{mem} = f_{mem}(\theta_{mem})$

\For{each vertex $v \in V(\workload)$}
    \State $t_{comp}(v) = \frac{\text{ComputeOps}(v)}{BW_{comp}}$
    \State $t_{mem}(v) = \frac{\text{MemoryOps}(v)}{BW_{mem}}$
    \State $t_{exec}(v) = \max(t_{comp}(v), t_{mem}(v))$
    \State $t_{idle}(v) = |t_{comp}(v) - t_{mem}(v)|$
    
    \If{$t_{comp}(v) > t_{mem}(v)$} \Comment{Memory subsystem idle}
        \State $\mathcal{I}_{log} = \mathcal{I}_{log} \cup \{(v, \text{memory\_idle}, t_{idle}(v))\}$
        \State $\frac{\partial \mathcal{L}}{\partial \theta_{mem}} += \alpha \cdot t_{idle}(v) \cdot \frac{\partial BW_{mem}}{\partial \theta_{mem}}$
    \ElsIf{$t_{mem}(v) > t_{comp}(v)$} \Comment{Compute subsystem idle}
        \State $\mathcal{I}_{log} = \mathcal{I}_{log} \cup \{(v, \text{compute\_idle}, t_{idle}(v))\}$
        \State $\frac{\partial \mathcal{L}}{\partial \theta_{comp}} += \alpha \cdot t_{idle}(v) \cdot \frac{\partial BW_{comp}}{\partial \theta_{comp}}$
    \EndIf
    
    \State $\nabla_{\theta} += \textsc{AccumulateParameterGradients}(v, t_{idle}(v), \mathcal{I}_{log})$
\EndFor

\State \textbf{return} $\nabla_{\theta}$, $\mathcal{I}_{log}$
\EndFunction

\Function{AccumulateParameterGradients}{$v$, $t_{idle}$, $\mathcal{I}_{log}$}
\State $\nabla_{local} = \mathbf{0}$
\For{each parameter $\theta_i$ affecting bandwidth}
    \State $\nabla_{local}[\theta_i] = t_{idle} \cdot \frac{\partial \text{PerformanceMetric}}{\partial \theta_i}$
\EndFor
\State \textbf{return} $\nabla_{local}$
\EndFunction
\end{algorithmic}
\end{algorithm}

\begin{algorithm}
\caption{Differentiable Hardware-Software Co-optimization}
\label{alg:co_optimization}
\footnotesize
\begin{algorithmic}[1]
\Function{OptimizeHardwareParameters}{$\workload$, $\theta_{init}$, $\mathcal{L}_{obj}$}
\State $\theta = \theta_{init}$, $t = 0$, $\nabla_{momentum} = \mathbf{0}$
\State $\beta = 0.9$, $\eta = 0.01$ \Comment{Momentum and learning rate}

\While{$t < T_{max}$ \textbf{and} $\|\nabla_{momentum}\|_2 > \epsilon_{conv}$}
    \State \textbf{Forward Pass:}
    \State $\conchwmodel = \textsc{InstantiateHardware}(\theta)$
    \State $\tuple{T_{exec}, E_{exec}, \mathcal{I}_{log}} = \textsc{SimulateWithIdleTracking}(\workload, \conchwmodel)$
    \State $\mathcal{L} = \mathcal{L}_{obj}(T_{exec}, E_{exec})$
    
    \State \textbf{Backward Pass:}
    \State $\nabla_{\theta} = \textsc{ComputeIdleTimeGradients}(\workload, \conchwmodel, \theta)$
    \State $\nabla_{momentum} = \beta \cdot \nabla_{momentum} + (1-\beta) \cdot \nabla_{\theta}$
    
    \State \textbf{Parameter Update with Constraints:}
    \State $\theta_{new} = \theta - \eta \cdot \nabla_{momentum}$
    \State $\theta = \textsc{ProjectToPhysicalConstraints}(\theta_{new})$
    
    \State \textbf{Convergence Check:}
    \If{$\textsc{BandwidthBalance}(\theta) < \epsilon_{balance}$}
        \State \textbf{break} \Comment{Achieved bandwidth balance}
    \EndIf
    \State $t = t + 1$
\EndWhile

\State \textbf{return} $\theta$, $\mathcal{I}_{log}$
\EndFunction

\Function{BandwidthBalance}{$\theta$}
\State $BW_{comp} = f_{comp}(\theta_{comp})$, $BW_{mem} = f_{mem}(\theta_{mem})$
\State \textbf{return} $\frac{|BW_{comp} - BW_{mem}|}{\max(BW_{comp}, BW_{mem})}$
\EndFunction
\end{algorithmic}
\end{algorithm}

\begin{algorithm}
\caption{Bandwidth-Balanced Scheduling with Idle Time Logging}
\label{alg:bandwidth_scheduling}
\footnotesize
\begin{algorithmic}[1]
\Function{ScheduleWithBandwidthAwareness}{$\mathcal{G}(V,E)$, $\conchwmodel$}
\State Initialize $\mathcal{S} = \emptyset$ \Comment{Schedule}
\State Initialize $\mathcal{I}_{detailed} = \emptyset$ \Comment{Detailed idle time log}
\State $\mathcal{R}_{mem} = \{\}$, $\mathcal{R}_{comp} = \{\}$ \Comment{Resource availability}

\For{each time step $t = 0, 1, 2, \ldots$}
    \State $V_{ready} = \{v \in V : \text{predecessors of } v \text{ completed}\}$
    
    \For{each $v \in V_{ready}$}
        \State $BW_{req,mem}(v) = \text{EstimateMemoryBandwidth}(v)$
        \State $BW_{req,comp}(v) = \text{EstimateComputeBandwidth}(v)$
        \State $BW_{avail,mem} = \conchwmodel(\text{memory\_bandwidth}) - \mathcal{R}_{mem}[t]$
        \State $BW_{avail,comp} = \conchwmodel(\text{compute\_bandwidth}) - \mathcal{R}_{comp}[t]$
        
        \If{$BW_{req,mem}(v) \leq BW_{avail,mem}$ \textbf{and} $BW_{req,comp}(v) \leq BW_{avail,comp}$}
            \State $\mathcal{S} = \mathcal{S} \cup \{(v, t)\}$ \Comment{Schedule vertex at time t}
            \State $\mathcal{R}_{mem}[t] += BW_{req,mem}(v)$
            \State $\mathcal{R}_{comp}[t] += BW_{req,comp}(v)$
            
            \State \textbf{Log Bandwidth Utilization:}
            \State $\rho_{mem} = \mathcal{R}_{mem}[t] / \conchwmodel(\text{memory\_bandwidth})$
            \State $\rho_{comp} = \mathcal{R}_{comp}[t] / \conchwmodel(\text{compute\_bandwidth})$
            \State $\mathcal{I}_{detailed} = \mathcal{I}_{detailed} \cup \{(v, t, \rho_{mem}, \rho_{comp})\}$
            
            \If{$|\rho_{mem} - \rho_{comp}| > \epsilon_{imbalance}$}
                \State $\textsc{LogBandwidthImbalance}(v, t, \rho_{mem}, \rho_{comp})$
            \EndIf
        \EndIf
    \EndFor
\EndFor

\State \textbf{return} $\mathcal{S}$, $\mathcal{I}_{detailed}$
\EndFunction
\end{algorithmic}
\end{algorithm}

\section{Appendix B: Mapping, Scheduling and Synthesis Details}

\subsection{Adaptive Prefetching Algorithm}

We present a bandwidth-aware prefetching algorithm that dynamically manages memory hierarchy utilization to optimize system performance. The algorithm maintains resource utilization below critical thresholds while maximizing data locality.

\begin{algorithm}
\caption{Bandwidth-Aware Adaptive Prefetching}
\label{alg:adaptive_prefetch}
\footnotesize
\begin{algorithmic}[1]
\State \textbf{Input:} Sorted node list $\mathcal{N}$, bandwidth threshold $\tau_{bw} = 0.9$, size threshold $\tau_{size} = 0.9$
\State \textbf{Output:} Optimized execution schedule with prefetch decisions
\State Initialize $\mathcal{D} \leftarrow \textsc{TopologicalSort}(\mathcal{N})$
\For{each node $n \in \mathcal{D}$}
    \State $\rho_{bw} \leftarrow \textsc{GetBandwidthUtilization}(n)$
    \State $\rho_{size} \leftarrow \textsc{GetSizeUtilization}(n)$
    \If{$\rho_{bw} > \tau_{bw}$}
        \State \textbf{continue} \Comment{Skip to avoid bandwidth saturation}
    \EndIf
    \If{$\rho_{size} > \tau_{size} \land \rho_{bw} < \tau_{bw}$}
        \State $\textsc{SetExecutionMode}(n, \text{streaming})$ \Comment{Enable memory streaming}
    \ElsIf{$\rho_{size} < \tau_{size} \land \rho_{bw} < \tau_{bw}$}
        \State $\textsc{PrefetchData}(n.\text{successor})$ \Comment{Proactive data prefetching}
    \EndIf
\EndFor
\end{algorithmic}
\end{algorithm}

\subsection{Performance-Oriented Mapping Strategies}

\subsubsection{Execution Time Optimization}

For latency-critical applications, we employ a multi-tiered optimization strategy that combines prefetching with adaptive tiling. The objective function minimizes the critical path length while maintaining resource constraints:

\begin{equation}
\min_{\mathcal{S}} \max_{p \in \mathcal{P}} \sum_{v \in p} t_{\text{exec}}(v, \mathcal{S})
\end{equation}

where $\mathcal{P}$ represents the set of all execution paths, $\mathcal{S}$ denotes the scheduling strategy, and $t_{\text{exec}}(v, \mathcal{S})$ is the execution time of vertex $v$ under schedule $\mathcal{S}$.

\subsubsection{Energy Efficiency Optimization}

For energy-constrained scenarios, we formulate the optimization as a multi-objective problem that balances computational efficiency with memory access patterns:

\begin{equation}
\min_{\mathcal{M}} \left( E_{\text{compute}}(\mathcal{M}) + E_{\text{memory}}(\mathcal{M}) + \lambda \cdot T_{\text{exec}}(\mathcal{M}) \right)
\end{equation}

where $\mathcal{M}$ represents the mapping configuration, $\lambda$ is the time-energy trade-off parameter, and energy components are defined as:

\begin{align}
E_{\text{compute}}(\mathcal{M}) &= \sum_{v \in V} \alpha_v \cdot \text{ops}(v) \cdot e_{\text{op}} \\
E_{\text{memory}}(\mathcal{M}) &= \sum_{v \in V} \left( \beta_v \cdot \text{reads}(v) \cdot e_{\text{read}} + \gamma_v \cdot \text{writes}(v) \cdot e_{\text{write}} \right)
\end{align}

\subsection{Data Locality and Reuse Optimization}

\subsubsection{Spatial and Temporal Locality Analysis}

We systematically identify and exploit data reuse opportunities through comprehensive locality analysis. The framework distinguishes between two primary locality patterns:

\begin{itemize}
    \item \textbf{Spatial Locality}: Exploiting contiguous memory access patterns through intelligent buffer management and cache-conscious data layout optimization.
    
    \item \textbf{Temporal Locality}: Maximizing data reuse by maintaining frequently accessed data in high-bandwidth, low-latency memory tiers.
\end{itemize}

The locality benefit function quantifies the potential performance improvement:

\begin{equation}
\mathcal{B}_{\text{locality}}(v) = \frac{\text{access\_frequency}(v) \cdot \text{data\_size}(v)}{\text{memory\_hierarchy\_cost}(v)}
\end{equation}

\subsubsection{Stencil Computation Optimization}

For stencil-based computations, we employ analytical models to quantify data reuse potential. The reuse factor $\mathcal{R}$ for a stencil operation is computed as:

\begin{equation}
\mathcal{R} = \frac{\text{total\_operations}}{\text{unique\_memory\_accesses}}
\end{equation}

For matrix multiplication and linear algebra kernels, we derive closed-form expressions for optimal tile sizes that maximize cache utilization while minimizing memory traffic.

\subsubsection{Loop Transformation and Blocking}

We extend the loop blocking methodology from \cite{yang2020interstellar} to energy-efficient execution through gradient-based optimization of tiling parameters. The optimization problem is formulated as:

\begin{equation}
\{x^*, y^*, c^*, k^*\} = \arg\min_{x,y,c,k} \mathcal{E}_{\text{total}}(\text{loop\_tiling}(x,y,c,k))
\end{equation}

where the energy cost function $\mathcal{E}_{\text{total}}$ incorporates both computational and memory access energy:

\begin{equation}
\mathcal{E}_{\text{total}} = \sum_{i=1}^{n} \left( \mathcal{E}_{\text{compute}}^{(i)} + \mathcal{E}_{\text{memory}}^{(i)} \right)
\end{equation}

The tiling parameters $\{x_1, x_2, \ldots, x_n\}$ are optimized across $n$ memory hierarchy levels using gradient descent with momentum:

\begin{equation}
x_{t+1} = x_t - \eta \nabla_x \mathcal{E}_{\text{total}} + \mu (x_t - x_{t-1})
\end{equation}

\subsubsection{Computation Reuse through Memoization}

For applications with recurring computational patterns, we implement intelligent memoization strategies. Common sub-expression elimination is performed through dataflow graph analysis, identifying recurring computation regions (RCRs) using the methodology from \cite{conners1999compiler}.

The memoization benefit is quantified through the cost-benefit analysis:

\begin{equation}
\mathcal{B}_{\text{memo}}(r) = \text{frequency}(r) \cdot \text{compute\_cost}(r) - \text{storage\_cost}(r) - \text{lookup\_cost}(r)
\end{equation}

where $r$ represents a recurring computation region, and memoization is applied when $\mathcal{B}_{\text{memo}}(r) > \theta_{\text{memo}}$ for a predefined threshold $\theta_{\text{memo}}$.

\subsection{Intermediate Representation and Scheduling Policies}

\subsubsection{Multi-Level Intermediate Representation}

Our framework employs a hierarchical intermediate representation (IR) strategy to accommodate diverse workload characteristics:

\begin{itemize}
    \item \textbf{LLVM IR}: For C/C++ programs, we leverage the LLVM intermediate representation to extract control-flow and data-flow information with precise memory access patterns.
    
    \item \textbf{Abstract Syntax Trees (AST)}: Python-based workloads are processed through AST analysis, enabling dynamic optimization of high-level computational patterns.
    
    \item \textbf{Dataflow Graph (DFG)}: Both representations are unified into a canonical DFG format that serves as the primary optimization target.
\end{itemize}

\subsubsection{Hardware Synthesis and Resource Allocation}

The hardware synthesis process follows a systematic two-phase approach:

\textbf{Phase 1: Schedule Generation}
We formulate the scheduling problem as a resource-constrained optimization:

\begin{equation}
\min_{\mathcal{T}} \max_{v \in V} \mathcal{T}(v) \quad \text{subject to} \quad \sum_{v \in \mathcal{A}(t)} r_j(v) \leq R_j \quad \forall t, j
\end{equation}

where $\mathcal{T}(v)$ represents the completion time of vertex $v$, $\mathcal{A}(t)$ is the set of active vertices at time $t$, $r_j(v)$ is the resource requirement of type $j$ for vertex $v$, and $R_j$ is the available resource capacity of type $j$.

\textbf{Phase 2: Resource Binding and Allocation}
The allocation process employs interval graph-based algorithms for conflict resolution and resource binding. The methodology from \cite{kollig1997simultaneous} is extended with performance-aware cost functions:

\begin{equation}
\mathcal{C}_{\text{bind}}(v, r) = \alpha \cdot \text{latency}(v, r) + \beta \cdot \text{energy}(v, r) + \gamma \cdot \text{area}(r)
\end{equation}

where $v$ represents a computation vertex, $r$ denotes a hardware resource, and $\{\alpha, \beta, \gamma\}$ are objective-specific weighting parameters.

\subsubsection{Datapath Generation and Functional Unit Mapping}

The datapath synthesis process systematically maps high-level constructs to hardware primitives:

\begin{itemize}
    \item \textbf{Control Structures}: \texttt{ast.For}, \texttt{ast.If}, and \texttt{ast.While} constructs are mapped to finite state machines with optimized control logic.
    
    \item \textbf{Computational Kernels}: \texttt{ast.Func} nodes are analyzed for specialized hardware opportunities, including:
    \begin{itemize}
        \item Matrix multiplication units for dense linear algebra
        \item Stencil processors for structured grid computations  
        \item Vector processing units for SIMD operations
    \end{itemize}
    
    \item \textbf{Memory Subsystem}: Default allocation employs scratchpad SRAM with intelligent partitioning based on bandwidth-compute balance requirements.
\end{itemize}

\subsection{Performance-Objective-Driven Scheduling}

\subsubsection{Latency-Optimized Scheduling}

For latency-critical applications, we employ an exponential penalty function that balances execution time with area constraints:

\begin{equation}
\mathcal{F}_{\text{latency}} = T_{\text{exec}} \cdot e^{\kappa(A_{\text{used}} - A_{\text{budget}})}
\end{equation}

where $T_{\text{exec}}$ is the execution time, $A_{\text{used}}$ and $A_{\text{budget}}$ represent used and budgeted area respectively, and $\kappa$ controls the penalty severity.

The optimization parameters include:
\begin{itemize}
    \item \textbf{Loop Unrolling Factors}: Determined through gradient-based exploration of the latency-area trade-off space
    \item \textbf{Memory Partitioning}: Optimized to eliminate memory bottlenecks while respecting area constraints
    \item \textbf{Pipeline Depth}: Balanced between throughput improvement and control complexity
    \item \textbf{Prefetch Buffer Allocation}: Sized based on working set analysis and bandwidth requirements
\end{itemize}

\subsubsection{Energy-Optimized Scheduling}

Energy-constrained scenarios employ a similar exponential formulation with energy as the primary objective:

\begin{equation}
\mathcal{F}_{\text{energy}} = E_{\text{total}} \cdot e^{\kappa(A_{\text{used}} - A_{\text{budget}})}
\end{equation}

where $E_{\text{total}}$ encompasses both dynamic and leakage energy consumption:

\begin{equation}
E_{\text{total}} = E_{\text{dynamic}} + E_{\text{leakage}} = \sum_{v \in V} \left( \alpha_v C_v V^2 f + P_{\text{leak},v} T_{\text{exec}} \right)
\end{equation}

The energy optimization process considers:
\begin{itemize}
    \item \textbf{Voltage-Frequency Scaling}: Dynamic adjustment based on computational intensity
    \item \textbf{Power Gating}: Intelligent shutdown of unused functional units
    \item \textbf{Clock Gating}: Fine-grained control of clock distribution
    \item \textbf{Memory Hierarchy Optimization}: Minimizing high-energy off-chip accesses
\end{itemize}

\subsubsection{Gradient-Based Parameter Optimization}

Both latency and energy objectives employ gradient descent with adaptive learning rates:

\begin{align}
\theta_{t+1} &= \theta_t - \eta_t \nabla_{\theta} \mathcal{F}(\theta_t) \\
\eta_t &= \eta_0 \cdot \left(1 + \frac{t}{\tau}\right)^{-\nu}
\end{align}

where $\theta$ represents the optimization parameters (unrolling factors, partition sizes, etc.), $\eta_0$ is the initial learning rate, $\tau$ is the decay time constant, and $\nu$ controls the decay rate.

The gradient computation leverages automatic differentiation through the mapping process, enabling efficient exploration of the high-dimensional parameter space while maintaining convergence guarantees.

\section{Appendix C : Theoretical Foundation and Proofs}

\subsection{Theorem 1: Optimal Bandwidth Balance Condition}

\textbf{Theorem 1.} \textit{Under hardware resource constraints, the optimal system configuration achieves bandwidth balance between compute and memory subsystems, i.e., $\delta t_{idle} = 0$ at optimality.}

\textbf{Proof:}
Let $\mathcal{H}(\theta)$ denote a hardware configuration parameterized by $\theta = \{\theta_{comp}, \theta_{mem}\}$, subject to area constraint $A_{total} \geq A(\theta)$. Define the execution time for workload $\mathcal{W}$ as:
\begin{equation}
T_{exec}(\theta, \mathcal{W}) = \sum_{v \in V(\mathcal{W})} \max\left(t_{comp}(v, \theta_{comp}), t_{mem}(v, \theta_{mem})\right)
\end{equation}

The idle time for each vertex $v$ is:
\begin{equation}
t_{idle}(v, \theta) = |t_{comp}(v, \theta_{comp}) - t_{mem}(v, \theta_{mem})|
\end{equation}

Consider the constrained optimization problem:
\begin{equation}
\mathcal{L}(\theta, \lambda) = T_{exec}(\theta, \mathcal{W}) + \lambda(A(\theta) - A_{total})
\end{equation}

At optimality, applying the Karush-Kuhn-Tucker (KKT) conditions:
\begin{align}
\frac{\partial \mathcal{L}}{\partial \theta_{comp}} &= \sum_{v \in V} \frac{\partial t_{idle}(v)}{\partial \theta_{comp}} + \lambda \frac{\partial A(\theta)}{\partial \theta_{comp}} = 0\\
\frac{\partial \mathcal{L}}{\partial \theta_{mem}} &= \sum_{v \in V} \frac{\partial t_{idle}(v)}{\partial \theta_{mem}} + \lambda \frac{\partial A(\theta)}{\partial \theta_{mem}} = 0
\end{align}

Since $\frac{\partial t_{idle}(v)}{\partial \theta} = 0$ if and only if $t_{comp}(v) = t_{mem}(v)$, the optimal solution satisfies bandwidth balance condition: $\delta t_{idle} = 0$. $\square$

\subsection{Theorem 2: Gradient Computation via Idle Time Logging}

\textbf{Theorem 2.} \textit{The gradient of the performance objective with respect to hardware parameters can be computed as the weighted accumulation of idle time measurements.}

\textbf{Proof:}
For performance objective $\mathcal{P}(\theta) = f(T_{exec}(\theta), E_{exec}(\theta))$, we have:
\begin{equation}
\frac{\partial \mathcal{P}}{\partial \theta_i} = \frac{\partial f}{\partial T_{exec}} \cdot \frac{\partial T_{exec}}{\partial \theta_i} + \frac{\partial f}{\partial E_{exec}} \cdot \frac{\partial E_{exec}}{\partial \theta_i}
\end{equation}

Since $T_{exec} = \sum_{v} \max(t_{comp}(v), t_{mem}(v))$, the gradient becomes:
\begin{equation}
\frac{\partial T_{exec}}{\partial \theta_i} = \sum_{v \in V} \mathbb{I}[\text{bottleneck}(v, \theta_i)] \cdot \frac{\partial t_{\text{bottleneck}}(v)}{\partial \theta_i}
\end{equation}

where $\mathbb{I}[\text{bottleneck}(v, \theta_i)]$ is the indicator function for the bottleneck resource. The idle time $t_{idle}(v) = |t_{comp}(v) - t_{mem}(v)|$ directly provides the gradient information:

\begin{equation}
\frac{\partial \mathcal{P}}{\partial \theta_i} = \sum_{v \in V} \alpha_v \cdot t_{idle}(v) \cdot \frac{\partial BW_i}{\partial \theta_i}
\end{equation}

where $\alpha_v$ is the weighting factor based on the bottleneck analysis. $\square$

\subsection{Corollary 1: Bandwidth Balance Optimality}

\textbf{Corollary 1.} \textit{A hardware configuration is Pareto-optimal if and only if it achieves bandwidth balance across all critical execution paths.}

This follows directly from Theorem 1, as any imbalance indicates potential for improvement without violating constraints.

\subsection{Example: Bandwidth-Balanced Matrix Multiplication}

Consider a tiled matrix multiplication workload with input matrices of size $N \times N$, mapped onto hardware with memory buffer size $B$ and $P$ parallel multipliers.

\textbf{Problem Formulation:}
\begin{align}
t_{mem}(tile) &= \frac{2B^2}{BW_{mem}} \quad \text{(read two } B \times B \text{ blocks)}\\
t_{comp}(tile) &= \frac{B^2}{P \cdot f_{clock}} \quad \text{(perform } B^2 \text{ multiply-adds)}\\
\text{Area constraint:} &\quad A_{SRAM}(B) + P \cdot A_{mult} \leq A_{total}
\end{align}

\textbf{Bandwidth Balance Condition:}
For optimal performance, we require $t_{mem} = t_{comp}$:
\begin{equation}
\frac{2B^2}{BW_{mem}} = \frac{B^2}{P \cdot f_{clock}}
\end{equation}

Solving for the optimal relationship:
\begin{equation}
P = \frac{BW_{mem}}{2 \cdot f_{clock}}
\end{equation}

\textbf{Gradient-Based Optimization:}
When bandwidth imbalance occurs, the idle time gradients guide parameter updates:
\begin{align}
\Delta B &\propto -\alpha \cdot t_{idle} \cdot \text{sign}(t_{comp} - t_{mem})\\
\Delta P &\propto -\alpha \cdot t_{idle} \cdot \text{sign}(t_{mem} - t_{comp})
\end{align}

This systematic approach ensures convergence to the bandwidth-balanced optimal configuration.

\subsection{Convergence Analysis}

\textbf{Theorem 3.} \textit{The gradient descent algorithm with idle time logging converges to a bandwidth-balanced configuration under Lipschitz smoothness assumptions.}

The proof follows standard convergence analysis for constrained optimization with the additional property that the gradient magnitude decreases as bandwidth balance is achieved, ensuring stable convergence.

\section{Appendix D : Software Class Diagram}

\begin{figure*}
    \centering
    \includegraphics[scale=0.7]{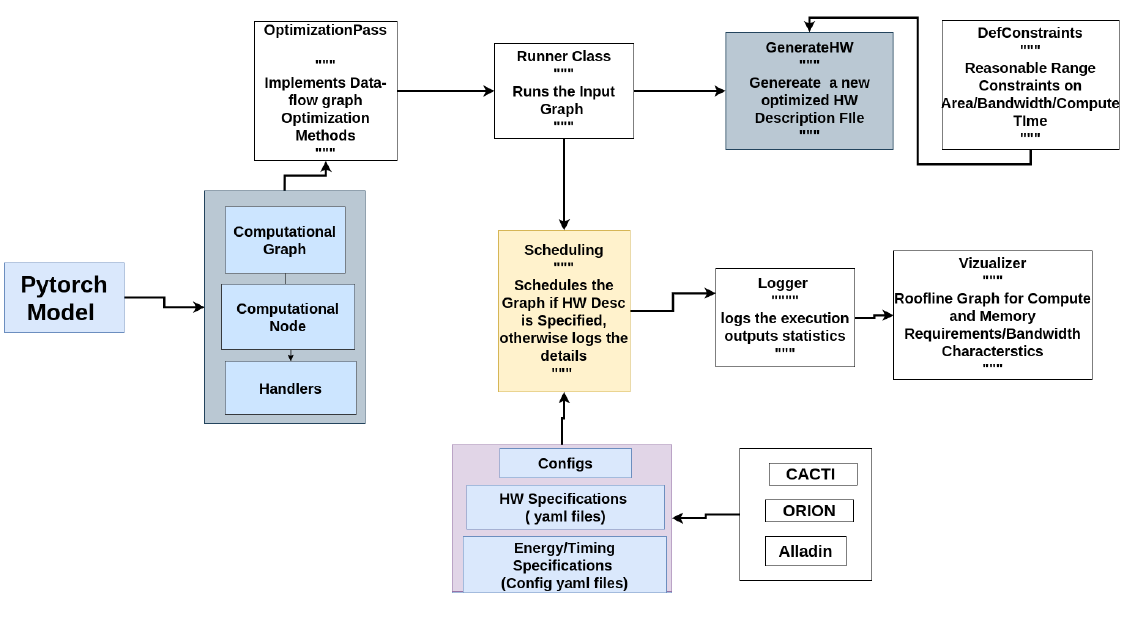}
    \caption{Software Class Diagram}
    \label{fig:acc}
\end{figure*}



\end{document}